# High-Fidelity Visual Structural Inspections through Transformers and Learnable Resizers

Kareem Eltouny[1†][0000-0003-3918-2297], Seyedomid Sajedi[1†][0000-0002-4552-4794], and Xiao Liang[1*][0000-0003-4788-8759]

[1] University at Buffalo, the State University of New York, Buffalo NY 14260, USA
*liangx@buffalo.edu
†Equal contribution

**Abstract.** Visual inspection is the predominant technique for evaluating the condition of civil infrastructure. The recent advances in unmanned aerial vehicles (UAVs) and artificial intelligence have made the visual inspections faster, safer, and more reliable. Camera-equipped UAVs are becoming the new standard in the industry by collecting massive amounts of visual data for human inspectors. Meanwhile, there has been significant research on autonomous visual inspections using deep learning algorithms, including semantic segmentation. While UAVs can capture high-resolution images of buildings' façades, high-resolution segmentation is extremely challenging due to the high computational memory demands. Typically, images are uniformly downsized at the price of losing fine local details. Contrarily, breaking the images into multiple smaller patches can cause a loss of global contextual information. We propose a hybrid strategy that can adapt to different inspections tasks by managing the global and local semantics trade-off. The framework comprises a compound, high-resolution deep learning architecture equipped with an attention-based segmentation model and learnable downsampler-upsampler modules designed for optimal efficiency and information retention. The framework also utilizes vision transformers on a grid of image crops aiming for high precision learning without downsizing. An augmented inference technique is used to boost the performance and reduce the possible loss of context due to grid cropping. Comprehensive experiments have been performed on 3D physics-based graphics models synthetic environments in the Quake City dataset. The proposed framework is evaluated using several metrics on three segmentation tasks: component type, component damage state, and global damage (crack, rebar, spalling).

## 1 Introduction

Innovations in sensing technology and deep learning have made a significant impact on autonomous structural visual inspections [1-3]. More recently, Camera-equipped Unmanned Aerial Vehicles (UAV) are proving to be an important asset in visual inspections by rapidly capturing high-resolution visual data of structures while increas-ing the safety of human workers. The massive amounts of collected footage are often post-processed by an off-the-shelf deep learning model. However, challenges arise when a trade-off between computational efficiency and high-fidelity precision needs to be achieved. In many situations, computational resource constraints necessitate the downsizing of high-resolution data for various tasks, including semantic segmentation, causing a loss in critical information.

Different tasks may need different levels of context in visual structural inspections. In this study, we propose a twin-model framework to tackle the challenges arising from the different natures of these tasks. **T**rainable **R**esizing for high-resolution **S**egmentation **Net**work (TRS-Net) is developed for component detection and damage severity segmentation, which highly rely on global context information. DmgFormer is dedicated to crack, rebar, and spalling segmentation, tasks that are highly sensitive to resolution loss. The proposed twin models are customized based on the latest advances in computer vision research on super-resolution, image segmentation, and transformer architectures.

## 2  Autonomous visual inspections framework

### 2.1  Components and damage state segmentation

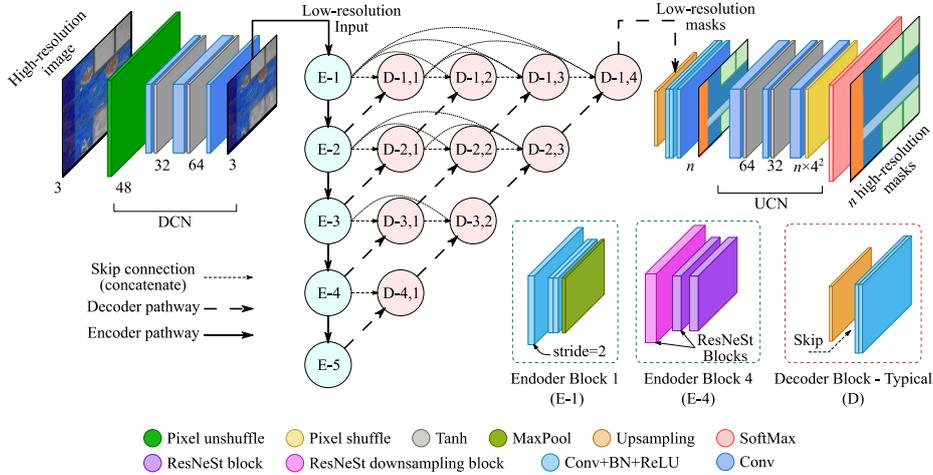

**Fig. 1.** TRS-Net architecture. (Conv: 2D convolution, BN: batch normalization, ReLU: rectified linear unit, E-$i$: encoder block, D-$i,j$: decoder block, n:classes).

It is a common practice to uniformly downsample high-resolution images before inserting them into deep learning segmentation models [4], and if needed, interpolation is used for upsampling the predicted masks. By giving equal significance to all pixels in the image, these operations may result in an unsatisfactory segmentation performance. In contrast, feeding high-resolution images directly to state-of-the-art segmen-tation networks is computationally inefficient and is usually met with GPU-memory constraints. To tackle these challenges, we propose high-resolution image segmentation based on learnable resizing. In this first part of the twin-model framework, we design a compound segmentation model with a two-layered encoder-decoder network trained end-to-end (Fig. 1). The outer-most layer includes efficient trainable downsampler and upsampler modules. These modules can be trained to focus on regions or pixels of high significance during both resizing processes. The internal encoder-decoder is a semantic segmentation model trained for low-resolution images and masks. The compound design has high memory efficiency. For example, for a single image forward and backward pass, a U-Net model with ResNet50 backbone needs 8.4 GB of GPU memory compared to 1.6 GB needed by TRS-Net.

Inspired by super-resolution networks [5], the upsampling convolutional network (UCN) is composed of three single-stride convolution layers followed by a pixel shuffle layer. The pixel shuffle layer reorganizes the low-resolution feature maps ($w/4$, $h/4$, $n \times 4^2$) to form four-times upscaled masks ($w$, $h$, $n$) corresponding to the number of prediction classes ($n$). The downsampling convolutional network (DCN) is almost an inverted version of UCN. It begins with a pixel unshuffle layer, a reverse operation to pixel shuffle, then followed by three convolutions. The pixel unshuffle operation guarantees that information is not lost during downsizing as pixels are only rearranged in the features axis. Fig. 2 presents a comparison between an example high-definition image (1920×1080), a four-times uniformly downsampled image (non-trainable), and our proposed trainable DCN. Compared to the original image, the uniformly downsampled image (Fig. 2b) shows a substantial loss of valuable information, particularly in the cracks and edges. DCN, on the other hand, can give more emphasis to such important details and retains them in a low-resolution form. In Fig. 2, for example, the cracks and edges appear thicker by DCN while the uniform downsampling method distorts both.

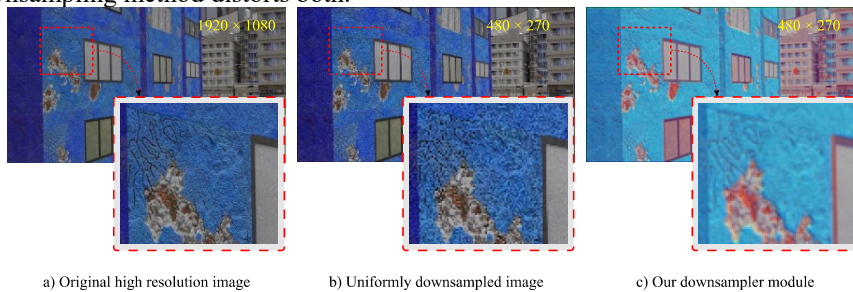

a) Original high resolution image     b) Uniformly downsampled image     c) Our downsampler module

**Fig. 2.** An example comparison between different downsampling techniques.

For the internal segmentation model, we use a ResNeSt50d backbone [6] with a U-Net++ decoder [7,8]. ResNeSt is a family of state-of-the-art image classification networks designed for efficiency and fast inference. ResNeSt introduced the split-attention mechanism which aims to capture the global context of the input by learning the channels' interdependencies in feature map groups. When compared to the traditional U-Net architecture, U-Net++ architecture adds complex dense convolutional blocks to the skip-connections between the encoder and decoder layers. These added dense blocks can bridge the semantic gap between the encoder and decoder, improving the gradient flow [6].

## 2.2 Damage segmentation

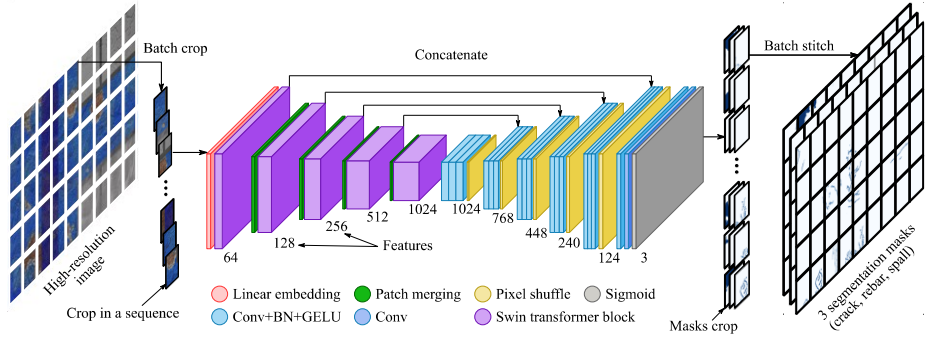

**Fig. 3.** DmgFormer architecture. (Conv: 2D convolution, BN: batch normalization, GELU: Gaussian error linear unit).

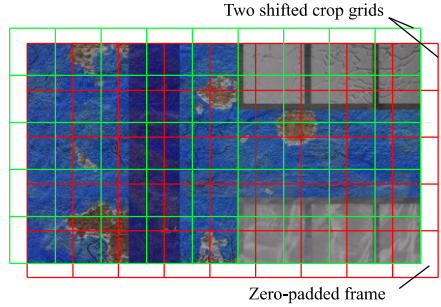

**Fig. 4.** Augmented inference in DmgFormer.

The downsampler and upsampler networks have significantly reduced the details lost in the resizing process. However, for segmenting fine damage patterns such as thin cracks and exposed rebars, we adopt an image cropping approach. We introduce DmgFormer for damage segmentation, a customized transformer model that relies on a grid-cropping mechanism (Fig. 3). In this method, the high-resolution image is first split into a grid of low-resolution crops, each passed to the segmentation model. Then, the high-resolution mask is reassembled from the low-resolution mask crops. Unlike components segmentation, our experiments showed that the loss in the global context is insignificant for the damage segmentation task. Yet, the cropping approach allows for precise high-fidelity damage segmentation while keeping practical computational expenses for training deep neural networks. As an example, a high-definition image (1920×1080) can be cropped into a 5×9 grid of 224×224 images after zero-padding the image for compatibility. During training, we add small random translation shifts to each crop centroid as an augmentation measure.

DmgFormer is equipped with a transformer backbone architecture. Transformers, originally used in natural language processing [9], are currently dominating many computer vision benchmarks including ImageNet and ADE20K [10-12]. Typical transformer networks, however, are difficult to implement in a U-Net-like segmentation architecture. Swin Transformer [13] tackles this challenge by building hierarchical feature maps using Patch Merging as the network gets deeper. Due to the locality of the self-attention mechanism, the network has linear computational complexity. We modified the backbone by increasing the patch size to 2 instead of 4 and having 5 Swin Transformer blocks. These adjustments allow for higher precision segmentation for tasks with fine masks such as crack damage. We also design 5 decoder Convolutional blocks that are compatible with the encoder to form the segmentation model. Additionally, and to increase confidence in predictions, we propose augmented inference (Fig. 4). In the inference phase, there are three types of zero padding for each direction which are right/top, left/bottom, and middle. This creates 8 different sets of grid crops. By feeding the same input image with different paddings, we get a set of predictions for each pixel. Then, the average is used as the result.

## 3     Case Study: QuakeCity Dataset

The QuakeCity dataset, which was released as part of the 2$^{nd}$ International Competition for Structural Health Monitoring [14], contains simulated UAV-captured images of buildings that have suffered earthquake-induced damage. The surface damage textures are produced using physics-based graphics models (PBGM) [15]. Each 1920×1080 RGB image is associated with five masks including components, components damage states, cracks, spalling, and exposed rebar. The components segmentation includes 7 different labels such as walls, beams, and window frames. The damage state segmentation includes 4 labels representing damage severity starting at "No damage" and up to "Severe damage". The dataset contains 3805 and 1004 images for training and testing respectively. At the time of writing, the annotations of the testing set have not been released. Therefore, we base our evaluation herein on a subset of the training dataset. We split the training set into 80%, 10%, and 10% subsets representing the training, validation, and testing set from this point forward.

All models are built and trained using the PyTorch library [16] and the University at Buffalo's Center for Computational Research [17]. In training, we implemented data augmentation techniques including a variety of random color manipulations and image transforms, which helped to reduce overfitting and increase the quality of predictions. Adam optimizer [18] was used for both models with learning rate schedulers having maximum learning rates of 1E-3 and 2E-4 for TRS-Net and DmgFormer, respectively. We used focal loss [19] as the loss function based on several experiments. The two models were trained for 300 epochs for each task. The internal segmentation model of TRS-Net was also trained for cracks, spalling, and exposed rebars segmentation using a crop size of 480×270 for comparison with DmgFormer.

## 4 RESULTS

The first part is evaluating TRS-Net for high-resolution components and damage state segmentation tasks. We compare our proposal with two other models. The first is the internal network of TRS-Net trained and tested on uniformly downsized images and masks. In the second model, we test the internal network on high-resolution images and masks by attaching non-trainable uniform upsampler and downsampler to the network's head and stem. We use precision (P), recall (R), F1-score (F1), and intersection-over-union (IoU) as evaluation metrics for all models. Table 1 shows the mean class metrics test results for components and damage state segmentation tasks. The first observation is that when using the non-trainable uniform downsampler and upsampler layers to predict high-resolution masks, the prediction quality deteriorates compared to the low-resolution model. In contrast, high-resolution segmentation via TRS-Net provides the best prediction results compared to the other models.

**Table 1.** Mean testing performance metrics for components and damage state segmentation.

|  | Components segmentation | | | | Damage state segmentation | | | |
| --- | --- | --- | --- | --- | --- | --- | --- | --- |
|  | P | R | F1 | IoU | P | R | F1 | IoU |
| Internal TRS-Net (low resolution) | 99.40 | 99.53 | 99.47 | 98.94 | 97.42 | 97.31 | 97.36 | 94.93 |
| Internal TRS-Net + resizing | 97.08 | 96.72 | 96.89 | 94.10 | 96.47 | 95.50 | 95.97 | 92.39 |
| TRS-Net | 99.56 | 99.56 | 99.56 | 99.13 | 98.60 | 97.83 | 98.21 | 96.52 |

**Table 2.** Testing performance metrics for damage segmentation.

|  | Precision | | | Recall | | |
| --- | --- | --- | --- | --- | --- | --- |
|  | Crack | Rebar | Spall | Crack | Rebar | Spall |
| DmgFormer | 85.82 | 88.76 | 97.94 | 87.83 | 85.99 | 98.60 |
| DmgFormer (AI-4) | 86.25 | 89.84 | 98.06 | 87.72 | 86.08 | 98.62 |
| DmgFormer (AI-8) | 86.41 | 89.97 | 98.10 | 87.70 | 86.27 | 98.63 |
| Internal TRS-Net | 82.69 | 88.01 | 96.72 | 87.17 | 85.01 | 97.35 |
|  | F1-score | | | IoU | | |
|  | Crack | Rebar | Spall | Crack | Rebar | Spall |
| DmgFormer | 86.82 | 87.35 | 98.27 | 76.70 | 77.54 | 96.60 |
| DmgFormer (AI-4) | 86.98 | 87.92 | 98.34 | 76.96 | 78.44 | 96.74 |
| DmgFormer (AI-8) | 87.05 | 88.08 | 98.37 | 77.07 | 78.70 | 96.78 |
| Internal TRS-Net | 84.87 | 86.48 | 97.03 | 73.72 | 76.18 | 94.24 |

Test results for cracks, spalling, and exposed rebars segmentation using DmgFormer and the internal network of TRS-Net are shown in Table 2. It is observed that by using 4 and 8 iterations of augmented inference (AI-4 & 8), prediction results improve but at the cost of inference speed. We also found in our experiment that a model trained to predict the three damage segmentation tasks concurrently offers improved prediction performance over three single models, each trained for one task. It is possible that the performance boost is because the model learns to differentiate between the three tasks in areas where they appear similar. An example of the framework mask predictions for all five tasks of the case study is depicted in Fig. 5.

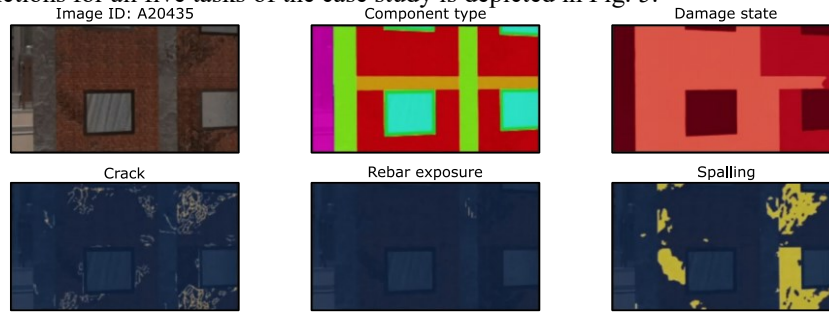

**Fig. 5.** Test prediction example from the QuakeCity dataset, image A20435 [15].

## 5 CONCLUSION

The rapid development of sensing technology and artificial intelligence has paved the way for safer, faster, and more accurate visual structural inspections. UAVs with their high level of autonomy can provide owners and inspectors with a massive amount of high-resolution frames of building façades after extreme events. While autonomous inspections prefer real-time and highly precise damage diagnosis, state-of-the-art deep learning vision models are getting increasingly complex and resource-hungry. In this study, we propose a twin segmentation models framework combining TRS-Net and DmgFormer which tackles these challenges using two philosophies. TRS-Net maintains the global context which is suitable for components and damage state segmentation tasks. TRS-Net is a memory and computationally efficient network owing to the information-preserving learnable downsampling and upsampling modules. In contrast, the transformer-equipped DmgFormer preserves the input resolution by using a grid-cropping mechanism which is a must for the highly sensitive damage segmentation tasks. Tested on the QuakeCity dataset, the twin-models framework proved to be a superior, yet efficient, segmentation tool that can be deployed for a variety of visual inspections tasks.